\documentclass[11pt]{article}

\usepackage[english]{babel}

\usepackage{hyperref}
\hypersetup{colorlinks=false}
\usepackage{color}
\usepackage{latexsym}
\usepackage{amssymb}
\usepackage{amsmath}
\usepackage{graphicx}
\usepackage{hhline}

\newtheorem{Proposition}{Proposition}[part]

\newtheorem{Lemma}{Lemma}[part]
\newtheorem{Corollary}{Corollary}[part]
\newtheorem{Remark}{Remark}[part]

\def\when{\; |\;}
\def \et{,\;}

\def \with{,\;\;\;}

\def \Sum{\displaystyle\sum}

\def \Frac{\displaystyle\frac}

\def \N{\mathbb{N}}
\def \R{\mathbb{R}}

\def \Z{\mathbb{Z}}
\def \E{\mathbb{E}}

\def \P{\mathbb{P}}

\def \Lc{{\cal L}}

 \def \Nc{{\cal N}}

\def \ni{\noindent}

\def \ep{\hbox{ }\hfill$\Box$}

\def\Ds#1{\Frac{\partial #1}{\partial s}}

\def\Dl#1{\Frac{\partial #1}{\partial \ell}}

\def\reff#1{{\rm(\ref{#1})}}

\def\beqs{\begin{eqnarray*}}
\def\enqs{\end{eqnarray*}}
\def\beq{\begin{eqnarray}}
\def\enq{\end{eqnarray}}

\begin{document}

\title{Semi Markov model  for market  microstructure}

\author{Pietro FODRA
             \\\small Laboratoire de Probabilit\'es et
             \\\small  Mod\`eles Al\'eatoires
             \\\small  CNRS, UMR 7599
             \\\small  Universit\'e Paris 7 Diderot
              \\\small and EXQIM
               \\\small  pietro.fodra91@gmail.com
             \and
              Huy\^en PHAM
             \\\small  Laboratoire de Probabilit\'es et
             \\\small  Mod\`eles Al\'eatoires
             \\\small  CNRS, UMR 7599
             \\\small  Universit\'e Paris 7 Diderot
             \\\small  pham@math.univ-paris-diderot.fr
             \\\small  CREST-ENSAE 
             \\\small and JVN Institute, Ho Chi Minh City
             }


\maketitle

\begin{abstract}
We introduce a new model for describing the fluctuations of a tick-by-tick single asset price. Our model is based on Markov renewal processes. 
We consider a point process associated to the timestamps of the price jumps, and marks associated to price increments. By modeling the marks with a suitable Markov chain, we can reproduce the strong mean-reversion of price returns known as microstructure noise. Moreover, by using Markov renewal processes, we  can model the presence of spikes in intensity of market activity, i.e. the volatility clustering, and consider dependence between price increments and jump times. 
We also provide simple parametric and nonparametric  statistical procedures for the estimation of  our model. 
We obtain closed-form formula  for the mean signature plot, and show the diffusive behavior of our model at large scale limit. 
We  illustrate our results by numerical simulations, and find  that our  model is consistent with empirical data on the Euribor future. 
\footnote{Tick-by-tick observation, from 10:00 to 14:00 during 2010, on the front future contract Euribor3m}
\end{abstract}

\vspace{3mm}


\newpage

\section{Introduction}
The modeling of tick-by-tick asset price  attracted a  growing  interest in the statistical and quantitative finance literature with  the availability of high frequency data. 
It  is  basically  split into two categories,  according to the philosophy guiding the modeling:

\noindent (i) The \textit{macro-to-microscopic}  (or econometric) approach, see e.g.  \cite{glojac01},  \cite{aitmykzha11}, 
\cite{RosembaumRobert2011},  interprets the observed price as a noisy representation of an unobserved one, typically assumed to be  a continuous It\^o semi-martingale: in this framework many important results exist on robust estimation of the realized volatility, but these models seem not tractable  for 
dealing with high frequency trading problems and stochastic control techniques, mainly because the state variables are latent rather than observed.

\noindent (ii) The \textit{micro-to-macroscopic}  approach (e.g. \cite{bauhau09}, \cite{bacetal10}, \cite{conlar10}, \cite{abejed11}, \cite{fautud12}) 
uses point processes, in particular Hawkes processes,  to describe the piecewise constant observed price, that moves on a discrete grid.  In contrast with the macro-to-microscopic approach, 
these models do not rely on the arguable existence assumption  of a fair or fundamental price, and focus on observable quantities, 
which makes the statistical estimation usually simpler.  Moreover, these models are able to reproduce several well-known stylized facts on high frequency data, see e.g. \cite{dac01}, \cite{boupot04}: 
\begin{itemize}
	\item \textit{Microstructure noise}: high-frequency returns are extremely anticorrelated, leading to a short-term mean reversion effect,  which is mechanically explicable by the structure of the limit order book. This effect manifests through an increase of  the realized volatility estimator (\textit{signature plot}) when the  observation frequency decreases from large to fine scales. 
	\item \textit{Volatility clustering}: markets alternates, independently of the intraday seasonality, between phases of high and low activity. 
	 \item At large scale, the price  process displays a {\it diffusion} behaviour.
\end{itemize}

In this paper, we aim  to provide a tractable model  of tick-by-tick asset  price  for liquid assets in a limit order book with a constant bid-ask spread,  in view of  application to  optimal high frequency trading problem, studied in the companion paper \cite{fodpha13b}.  
We start from a model-free description of the piecewise constant mid-price, i.e. the half of the  best bid and best ask price,  characterized by a marked point process $(T_n,J_n)_n$, where 
the timestamps $(T_n)$ represent the jump times of the asset price associated to a counting process $(N_t)$, and the marks $(J_n)$ are the price increments. 
We then use a Markov renewal process (MRP) for modeling  the marked point process. Markov renewal theory \cite{limopr01}  is largely studied in reliability for describing failure of systems and machines, and one purpose of this paper is to show how it can be applied for market microstructure.  By considering a suitable Markov chain modeling for 
$(J_n)$, we are able to reproduce the mean-reversion of price returns, while allowing  arbitrary jump size, i.e. the price can jump of more than one tick. 
 Furthermore,  the counting process $(N_t)$, which may depend on the Markov chain, models the volatility clustering, i.e. the presence of spikes in the volatility of the stock price, and we also preserve a Brownian motion behaviour of the price process at macroscopic scales. 
 Our MRP model is a rather simple but robust model, easy to understand, simulate and estimate, both parametrically and non-parametrically, based on i.i.d. sample data. 
An important feature of the MRP approach is the semi Markov property, meaning that the price process can be embedded into a Markov system with few additional and observable state variables. This will ensure the tractability of the model for  applications to market making  and statistical arbitrage.

The  outline of the paper is the following. In Section \ref{section:model}, we describe the MRP model for the asset mid-price, and provides statistical estimation procedures. We  show  the simu\-lation of the model, 
and the semi-Markov property of the price process.  We also discuss the comparison of our model with Hawkes processes used for modeling asset prices and microstructure noise.   
Section \ref{section:diffusive} studies the  diffusive limit  of the asset price at macroscopic scale.  In Section  \ref{section:signature}, we derive analytical formula for  the mean signature plot, and compare with real data.   Finally, Section 5 is devoted to a conclusion and  extensions for future research.

\section{Semi-Markov model}\label{section:model}

\setcounter{equation}{0} \setcounter{Assumption}{0}
\setcounter{Theorem}{0} \setcounter{Proposition}{0}
\setcounter{Corollary}{0} \setcounter{Lemma}{0}
\setcounter{Definition}{0} \setcounter{Remark}{0}

We describe  the tick-by-tick  fluctuation of a univariate stock price by means of a marked point process $(T_n,J_n)_{n\in\N}$. 
The increasing sequence  $(T_n)_n$  represents the jump (tick) times of the asset price, while the marks sequence  $(J_n)_n$ valued in 
the finite set 
$$E=\{+1,-1,\ldots,+m,-m\} \subset \Z\setminus\{0\}$$ 
represents the price increments.  Positive (resp. negative) mark means that price jumps upwards (resp. downwards). The continuous-time price process is a piecewise constant, pure jump process,  given by
\beq \label{price}
P_t &=& P_0 +  \sum_{n=1}^{N_t} J_n, \;\;\; t \geq 0,
\enq
where $(N_t)$ is the counting process associated to the tick  times $(T_n)_n$, i.e. 
\beqs
N_t &=& \inf\Big\{n:  \sum_{k=1}^n T_k \leq t \Big\}
\enqs
Here, we normalized the tick size (the minimum variation of the price) to $1$, and the asset price $P$  is considered e.g. as the last quotation for the mid-price, i.e. the mean between the best-bid and best-ask price.  Let us mention that the continuous time dynamics  \reff{price} is a model-free description of a piecewise-constant price process in a market microstructure. We decouple the modeling on one hand of the clustering of  trading activity via the point process $(N_t)$, and on the other hand of the microstructure noise (mean-reversion of price return) via the random marks $(J_n)_n$.

\subsection{Price return modeling}

We write the price return as
\beq \label{Jn} 
J_n &=& \hat J_n\,\xi_n,  \;\;\; n \geq 1,
\enq
where $\hat J_n$ $:=$  sign$(J_n)$ valued in $\{+1,-1\}$ indicates 
whether the price jumps upwards or downwards, and $\xi_n$ $:=$ 
$|J_n|$ is the absolute size of the price increment. We consider  that dependence of the price returns occurs  through their direction, and we assume that only the current price direction will impact the next  jump direction. Moreover, we assume  that the  absolute size of the price returns are independent and also independent of the direction of the jumps. Formally, this means that
\begin{itemize}
\item $(\hat J_n)_n$  is an irreducible  Markov chain with probability transition matrix
\beq \label{hatQmatrix}
\hat Q &=& \left( \begin{array}{cc}
			   \frac{1+\alpha_+}{2} & \frac{1-\alpha_+}{2} \\
			   \frac{1- \alpha_-}{2} & \frac{1 + \alpha_-}{2} 
			   \end{array}
		  \right) 
\enq
with $\alpha_+,\alpha_-$ $\in$ $[-1,1)$. 
\item $(\xi_n)_n$ is an i.i.d.  sequence valued in $\{1,\ldots,m\}$, independent of $(\hat J_n)$, with distribution law: 
$p_i$ $=$ $\P[\xi_n = i]$ $\in$ $(0,1)$, $i$ $=$ $1,\ldots,m$. 
\end{itemize}
In this case, $(J_n)$ is an irreducible  Markov chain with  probability transition matrix given by:
\beq \label{Qparti}
Q &=& \left( \begin{array}{ccc}
		    p_1 \hat Q & \ldots & p_m \hat Q \\
		    \vdots & \ddots & \vdots \\
		    p_1\hat Q & \ldots & p_m\hat Q
		    \end{array}
	 \right).
\enq
We could model in general $(J_n)_n$ as a Markov chain with transition matrix $Q$ $=$ $(q_{ij})$ involving $2m(2m-1)$ parameters, while  under the above assumption, the matrix $Q$ in \reff{Qparti} involves only $m+1$ parameters to be estimated.  Actually, on real data, we often observe that the number of consecutive downward and consecutive upward jumps are roughly equal. We shall then consider the {\it symmetric case} where
\beq \label{sym}
\alpha_+ &=& \alpha_-  \; =:  \; \alpha.  
\enq
In this case, we have a nice  interpretation of the parameter $\alpha$ $\in$ $[-1,1)$.

\begin{Lemma} \label{lemalpha}
In the symmetric case, the invariant distribution of  the Markov chain $(\hat J_n)_n$  is  
$\hat\pi$ $=$ $(\frac{1}{2},\frac{1}{2})$, and the invariant distribution of   $(J_n)_n$   is $\pi$ $=$ $(p_1\hat\pi, \ldots,p_m\hat\pi)$. Moreoever,   we have: 
\beq \label{correlalpha}
\alpha &=& 
{\rm corr}_{\pi}\big(\hat J_n,\hat J_{n-1}\big), \;\;\; \forall n \geq 1,
\enq
where ${\rm corr}_{\pi}$ denotes the correlation under the stationary probability $\P_\pi$ starting from the initial distribution $\pi$. 
\end{Lemma}
\ni {\bf Proof.}  We easily check that under the symmetric case,  $\hat\pi \hat Q$ $=$ $\hat\pi$ for $\hat\pi$ $=$ $(1/2,1/2)$, which means that $\hat\pi$ is the invariant distribution  of  the Markov chain $(\hat J_n)_n$. Consequently, $\pi$ $=$ $(p_1\hat\pi, \ldots,p_m\hat\pi)$ satisfies $\pi Q$ $=$ $\pi$, i.e. $\pi$ is the invariant distribution of $(J_n)_n$, and so  under $\P_\pi$, $(\hat J_n)_n$ (resp. $(J_n)_n$) is distributed according to $\hat\pi$ (resp. $\pi$). Therefore, 
$\E_\pi[\hat J_n]$ $=$ $0$ and $\mbox{Var}_\pi[\hat J_n]$ $=$ $1$.  We also have for all $n$ $\geq$ $1$, by definition of $\hat Q$:  
\beqs
\E_\pi\big[\hat J_n \hat J_{n-1}\big] &=& \E_\pi \Big[  \frac{1+\alpha}{2} \big(\hat J_{n-1} \big)^2  
- \frac{1- \alpha}{2}\big( \hat J_{n-1} \big)^2 \Big]  \; = \; \alpha, 
\enqs
which  proves the relation \reff{correlalpha}  for $\alpha$. 
\ep

 \vspace{2mm}

Lemma \ref{lemalpha}  provides a direct interpretation of the parameter $\alpha$ as the correlation between two consecutive price return  directions.  
The case $\alpha$ $=$ $0$ means that price returns are independent,  while $\alpha$ $<$ $0$ (resp. $\alpha$ $>$ $0$) corresponds to a mean-reversion (resp. trend) of price return.



\vspace{2mm}

We also have another equivalent formulation of the Markov chain,  whose proof is trivial and left to the reader.

\begin{Lemma} \label{lember}
In the symmetric case,  the Markov chain $(\hat J_n)_n$ can be written as:
\beqs \label{signJn}
\hat J_n &=& \hat J_{n-1} B_n, \;\;\; n \geq 1,
\enqs
where $(B_n)_n$ is a sequence of i.i.d. random variables  with Bernoulli distribution  on $\{+1,-1\}$, and parameter $(1+\alpha)/2$, i.e. of mean 
$\E[B_n]$ $=$ $\alpha$.  The price increment Markov chain $(J_n)_n$ can also be written  in an explicit induction form as:
\beq \label{inducJn}
J_n &=& \hat J_{n-1} \zeta_n,
\enq
where $(\zeta_n)_n$ is a sequence of i.i.d. random variables valued in $E$ $=$ $\{+1,-1,\ldots,+m,-m\}$, and with distribution: 
$\P[\zeta_n=k]$ $=$ $p_k(1+ \mbox{sign}(k)\alpha)/2$. 
\end{Lemma}

\vspace{2mm}

The  above Lemma is useful for an  efficient estimation of $\alpha$. Actually,  by the strong law of large numbers, we have  a consistent estimator of $\alpha$: 
\beqs
\hat\alpha^{(n)} &=& \frac{1}{n} \sum_{k=1}^n \frac{\hat J_k}{\hat J_{k-1}} \; = \; \frac{1}{n} \sum_{k=1}^n 
\hat J_k \hat J_{k-1}. 
\enqs
The variance of this estimator is known, equal to $1/n$, so that 
this estimator is efficient and we have a confidence interval from the central limit theorem: 
\beqs
\sqrt{n}(\hat\alpha^{(n)} - \alpha) & \stackrel{ (d) }{\longrightarrow} & \Nc(0,1), \;\;\; \mbox{ as } n \rightarrow \infty. 
\enqs
The estimated parameter for the chosen dataset is $\hat\alpha$ $=$ $-87.5\%$, 
which shows as expected a strong anticorrelation of price returns.  
In the case of several tick jumps $m$ $>$ $1$, the probability $p_i$ $=$ $\P[\xi_n=i]$ may be estimated from the classical empirical frequency: 
$$\hat p_i^{(n)}=\frac{1}{n} \sum_{k=1}^n 1_{\xi_k=i}\with i=1,\ldots,m$$

 \subsection{Tick times modeling}

 In order to describe  volatility clustering, we look for a counting process $(N_t)$ with an intensity increasing every time the price jumps, and decaying with time. We propose a modeling via Markov renewal process.  

\vspace{2mm}






 
Let us denote by $S_n$ $=$ $T_n -T_{n-1}$,  $n$ $\geq$ $1$,  the inter-arrival times associated to $(N_t)$. We assume that conditionally on the 
jump marks $(J_n)_n$,  $(S_n)_n$ is an independent sequence of positive random times, with distribution depending on the current and next jump mark: 
\beqs
F_{ij}(t) &=& \P[ S_{n+1} \leq t \when J_{n} = i, J_{n+1}=j], \;\;\; (i,j) \in E.
\enqs
We then say that $(T_n,J_n)_n$ is a Markov Renewal Process (MRP)  with transition kernel:
\beqs
\P[ J_{n+1}=j, S_{n+1} \leq t \when J_n=i] &=& q_{ij} F_{ij}(t), \;\;\;  (i,j) \in E. 
\enqs
In the particular case where $F_{ij}$ does not depend on $i,j$, the point process $(N_t)$ is independent of $(J_n)$, and called a renewal process, Moreover, if  $F$ is the exponential distribution, $N$ is a Poisson process.  Here, we allow in general dependency between jump marks and renewal times, and we refer to the symmetric case when $F_{ij}$ depends only on the sign of  $ij$, by setting: 
\beq \label{symF}
F_+(t) \; = \; F_{ij}(t), \;\;\;  \mbox{ if } \; ij > 0,  & &  F_-(t) \; = \; F_{ij}(t), \;\;\;  \mbox{ if } \; ij <  0. 
\enq
In other words, $F_+$ (resp. $F_-$) is the distribution function of inter-arrival times given two consecutive jumps in the same (resp. opposite) 
direction, called trend (resp. mean-reverting) case. 
Let us also introduce  the marked  hazard function:
\beq \label{hazard}
h_{ij}(t) &=& \lim_{\delta\downarrow 0} \frac{1}{\delta}  \P[  t \leq S_{n+1} \leq t + \delta \et J_{n+1} = j \when S_n \geq t \et  J_n   = i ], \;\;\;  t \geq 0, 
\enq
for $i,j$ $\in$ $E$, which represents the instantaneous probability that there will be a jump with mark $j$, given that there were no jump during the elapsed time $t$, and the current mark is $i$. By assuming that the distributions $F_{ij}$ of the renewal times $S_n$ admit a density $f_{ij}$,  
we may write $h_{ij}$ as: 
\beqs
h_{ij}(t) &=& q_{ij}  \frac{f_{ij}(t)}{1-H_i(t)}  =: q_{ij} \lambda_{ij}, 
\enqs
where 
\beqs
H_i(t) &=& \P[ S_{n+1}  \leq t \when J_n = i ] \; = \; \sum_{j\in E} q_{ij} F_{ij}(t),
\enqs
is the conditional distribution of the renewal time in state $i$.  In the symmetric case \reff{sym}, \reff{symF}, we have 
\beqs
h_{ij}(t) &=& p_j \Big( \frac{1+{\rm sign}(ij)\alpha}{2}\Big) \lambda_{{\rm sign}(ij)}(t),
\enqs
with jump intensity
\beq \label{jump+-}
\lambda_\pm(t) &=& \frac{f_\pm(t)}{1- F(t)},
\enq
where $f_\pm$ is the density of $F_\pm$, and
\beqs
F(t) &=& \frac{1+\alpha}{2} F_+(t) +  \frac{1- \alpha}{2} F_-(t). 
\enqs

Markov renewal processes   are used in many applications, especially in reliability. 
Cla\-ssical examples of  renewal distribution  functions are the ones corresponding to the  Gamma and Weibull distribution, with density given by:
\beqs
f_{Gam}(t) \; = \;  \frac{t^{\beta-1}e^{-t/\theta}}{\Gamma(\beta)\theta^\beta} 
& & \;\;\; f_{Wei}(t) \; = \; \frac{\beta}{\theta} \Big( \frac{t}{\theta} \Big)^{\beta-1} e^{-\beta t/\theta}
\enqs
where $\beta$ $>$ $0$, and $\theta$ $>$ $0$ are the shape and scale  parameters, and $\Gamma$ (resp. $\Gamma_t$) is  the  Gamma (resp.  lower incomplete Gamma) function: 
\beq \label{defgamma}
\Gamma(\beta) \; =  \;   \int_0^\infty s^{\beta-1} e^{-s} ds, \;\; 
& & \;\; \Gamma_t(\beta) \; = \;  \int_0^t s^{\beta-1} e^{-s} ds,
\enq


\vspace{1mm}

\subsection{Statistical procedures}

We design  simple statistical procedures for the estimation of the distribution and jump intensities of the renewal times.  Over an observation period, pick a subsample of i.i.d. data:
\beqs
\{ S_k = T_k - T_{k-1}: k \; s.t. \; J_{k-1} = i, \; J_k = j \},  
\enqs
and set:
\begin{eqnarray*}
I_{ij} & = & \#\{ k \;  s.t. \; J_{k-1} = i, \; J_k = j \} \\ 
I_{i} & = & \#\{ k \;  s.t. \; J_{k-1} = i \}, 
\end{eqnarray*} 
with cardinality respectively $n_{ij}$ and  $n_i$. In the symmetric case, we also denote by 
\beqs
I_{\pm} & = & \#\{ k \;  s.t. \; sign(J_{k-1}J_k) = \pm \},  
\enqs
with cardinality $n_\pm$.   We describe both parametric and nonparametric estimation.

\vspace{2mm}

\ni  $\bullet$ {\bf Parametric estimation}

\vspace{1mm}

\noindent We  discuss  the parametric estimation of the distribution $F_{ij}$ of the renewal times when considering Gamma or Weibull distributions with shape and scale parameters 
$\beta_{ij}$, $\theta_{ij}$.   We can indeed consider the Maximum Likelihood Estimator (MLE)  $(\hat\beta_{ij},\hat\theta_{ij})$, which are solution to the equations:  
\beqs
\ln\hat\beta_{ij}   - \frac{\Gamma'(\hat\beta_{ij})}{\Gamma(\hat\beta_{ij})} &=& \ln\Big( \frac{1}{n_{ij}} \sum_{k=1}^{n_{ij}} S_k \Big)  
- \frac{1}{N} \sum_{k=1}^{n_{ij}} \ln S_k \\
\hat \theta_{ij} &=&   \frac{1}{ \hat\beta_{ij}}  \bar S_{n_{ij}}, 
\;\;\; \mbox{ with }  \; \bar S_{n_{ij}} \; := \;  \frac{1}{n_{ij}} \sum_{k=1}^{n_{ij}}  S_k 
\enqs
There is no closed-form solution for $\hat\beta_{ij}$, which  can be obtained numerically  e.g. by Newton method. Alternatively, since the first two moments of the Gamma distribution $S$ $\leadsto$ $\Gamma(\beta,\theta)$ are explicitly given in terms of the shape and scale parameters, namely:
\beq \label{momentgam}
\beta \; = \;  \frac{\big(\E[S]\big)^2}{{\rm Var}[S]}, \;\;\; & & \;\;\;   \frac{1}{\theta} \; = \;  \frac{\E[S]}{{\rm Var}[S]},
\enq
we can estimate $\beta_{ij}$ and $\theta_{ij}$ by moment matching method, i.e. by replacing in \reff{momentgam} the mean and variance by their empirical estimators, which leads to:
\beqs
\tilde\beta_{ij} \; = \;  \frac{ n_{ij} \bar S_{n_{ij}}^2 } {\Sum_{k=1}^{n_{ij}} (S_k - \bar S_{n_{ij}})^2}, \;\;  & & \;\; 
\frac{1}{\tilde\theta_{ij}} \; = \; \frac{ n_{ij} \bar S_{n_{ij}} } { \Sum_{k=1}^{n_{ij}} (S_k - \bar S_{n_{ij}})^2}. 
\enqs

We performed  this parametric estimation method for the Euribor on the year 2010, from 10h to 14h,  with one tick jump, and obtain the following estimates  in Table \ref{tableparam}.

\begin{table}
\begin{center} 
\begin{tabular}{|cc|cc|}
\hline
i & j & shape & scale \\
\hline
+1 & +1 & 0.27651097 & 2187 \\
-1 & -1 & 0.2806104 & 2565.371   \\
+1 & -1 & 0.07442401& 1606.308 \\
-1 & +1 & 0.06840708 & 1508.155  \\
\hline
\end{tabular} 
\caption{Parameter estimates for the renewal times of a Gamma distribution}
\label{tableparam}
\end{center} 
\end{table}

\begin{figure}[h!] 
\centering
\includegraphics[width=9cm,height=9cm]{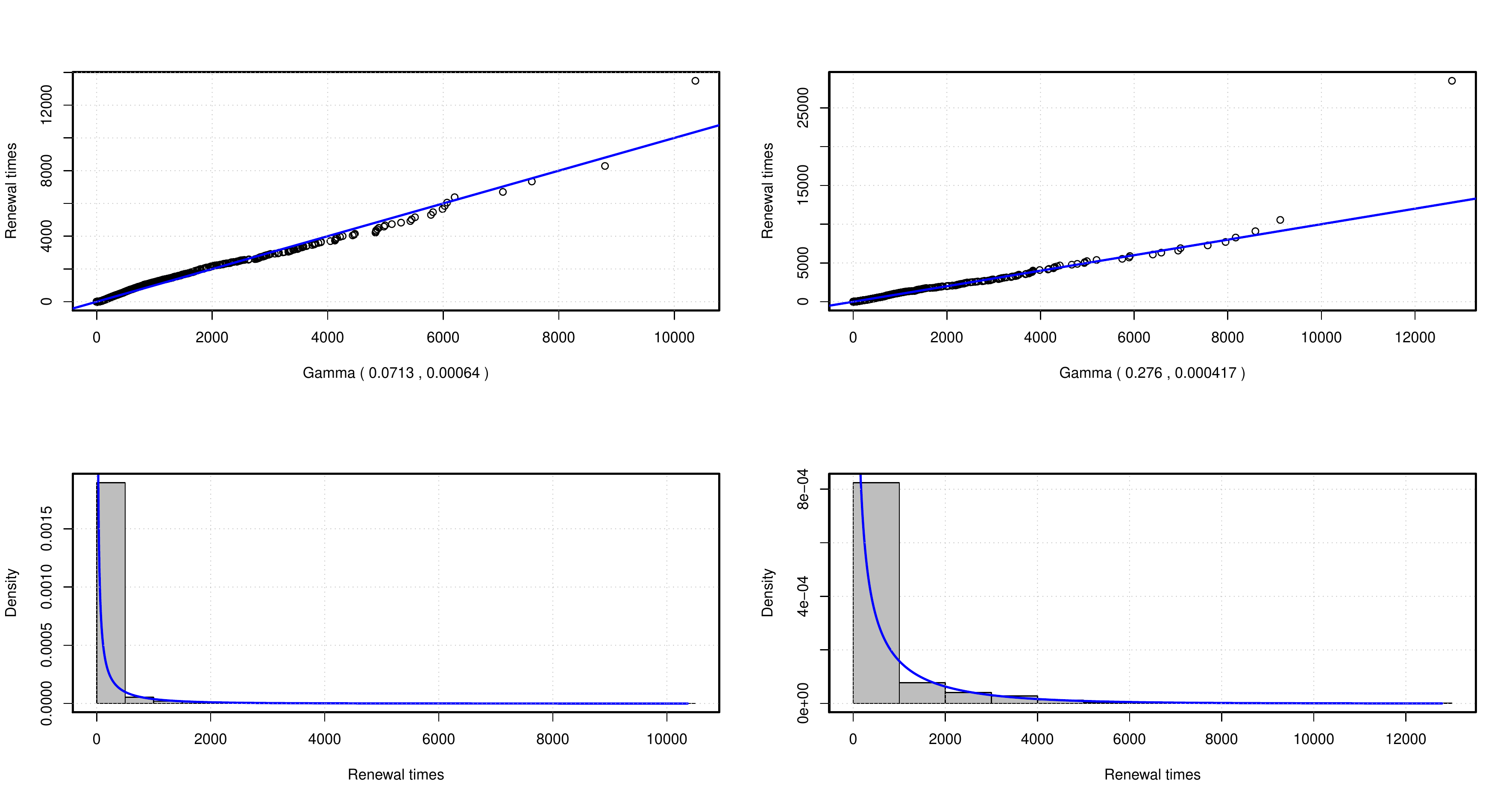} 
\caption{\small{QQ plot and histogram of $F_{-}$(left)  and $F_{+}$ (right). Euribor3m, 2010, 10h-14h.}}
\label{figplot}
\end{figure}

We observed that the shape and scale parameters depend on the product $ij$ rather than on  $i$ and $j$ separately. In other words, the distribution of the renewal times are symmetric in the sense of \reff{symF}.   Hence, we   performed again in Table  \ref{tabletrendmean}.  the  parametric estimation 
$(\tilde\beta_+,\tilde\theta_+)$ and $(\tilde\beta_-,\tilde\theta_-)$ for the shape and scale parameters 
by distinguishing only samples for $ij$ $=$ $1$ (the trend case) and $ij$ $=$ $-1$ (the mean-reverting case).  
We  also provide in Figure \ref{figplot} graphical tests  of the goodness-of-fit for our estimation results. The estimated value $\tilde\beta_+$, and 
$\tilde\beta_-$ $<$ $1$ for the shape parameters,  can be interpreted when considering the hazard rate function of $S_{n+1}$  given current and next marks:
\beqs
\hat\lambda_\pm(t) &:=& \lim_{\delta\downarrow 0} \frac{1}{\delta}  \P[  t \leq S_{n+1} \leq t + \delta | S_n \geq t,  {\rm sign}(J_n J_{n+1}) = \pm ] \\   
&=& \frac{f_{\pm}(t)}{1-F_\pm(t)}, \;\;\; t \geq 0, 
\enqs
when $F_\pm$ admits a density $f_\pm$ (Notice that $\hat\lambda_\pm$ differs from $\lambda_\pm$). 
$\hat\lambda_+(t)$ (resp. $\hat\lambda_-(t))$ is the instantaneous probability  of price jump given that there were no jump during an elapsed time $t$, and the current and next jump are in the same (resp. opposite) direction.  For the Gamma distribution with shape and scale parameters $(\beta,\theta)$, the hazard rate function is given by:
\beqs
\hat\lambda_{Gam}(t) & = &  \frac{1}{\theta} \frac{ \Big( \frac{t}{\theta} \Big)^{\beta-1} e^{-t/\theta}}{\Gamma(\beta)-\Gamma_{t/\theta}(\beta)}, \;\;\; 
\enqs
and is decreasing in time if and only if the shape parameter $\beta$ $<$ $1$, which means that the more the time passes, the less is the probability that an event occurs.  Therefore, the estimated values  $\hat\beta_+$, and $\tilde\beta_-$ $<$ $1$ are  consistent with the  volatility clustering:  
when a jump occurs, the probability of another jump in a short period is high, but if the event does not take place soon, then the price is likely to stabilize. Using this modeling, a long period of constant price corresponds to a renewal time in distribution tail. 
On the contrary, since renewal times are likely to be small when $\beta<1$, most of the jumps will be extremely close. 
We also notice that the parameters in the trend and mean-reverting case differ significantly.  This can be explained as follows: trends, i.e. two consecutive jumps in the same direction, are extremely rare (recall that $\alpha\approx -90\%$), since, in order to take place, market orders either have to clear two walls of liquidity or there must be a big number of cancellations. Since these events are caused by specific market dynamic, it is not surprising that their  renewal law differ   from the mean-reverting case.

\begin{table}
\begin{center} 
\begin{tabular}{|c|cc|}
\hline
i  j & shape & scale \\
\hline
+1  (\mbox{trend}) & 0.276225 & 2397.219   \\
-1 (\mbox{mean-reverting}) &  0.07132677 & 1561.593   \\
\hline
\end{tabular} 
\caption{Parameter estimates in the trend and mean-reverting case}
\label{tabletrendmean}
\end{center}  
\end{table}

\vspace{2mm}

\ni $\bullet$ {\bf Non parametric estimation} 

\vspace{1mm}

\noindent Since the renewal times form  an i.i.d. sequence, we are able to perform a non parametric estimation by kernel method of the density $f_{ij}$ and the jump intensity.  We  recall the smooth kernel method for the density estimation, and, in a similar way, we will give the one for $h_{\ij}$ and 
$\lambda_{ij}$.  Let us start with the empirical 
histogram of the density, which is constructed as follows. For every collection of breaks 
$$\{0<t_1<...<t_M\leq\infty\}  \with \delta_r \; := \; t_{r+1}-t_r$$ 
we  bin the sample $(S_k=T_k-T_{k-1})_{k=1,\ldots,n}$ and define the empirical histogram of $f_{ij}$ as:
\begin{eqnarray*}
f^{hst}_{ij}\,(t_r)
&=& \frac{1}{\delta_r} \frac{\#\{k \in I_{ij} \when t_r\leq S_k< t_{r+1}\}}{n_{ij}}. 
\end{eqnarray*}
By the strong law of large numbers, when $n_{ij}\rightarrow\infty$, this estimator converges to 
$$\frac{1}{\delta_r}\mathbb{P}\left[t_r\leq S_k<t_{r+1}\right],$$
which is a first order approximation of $f_{ij}(t_r)$. Anyway, this estimator depends on the choice of the binning, which has  not necessarily small (nor equal) $\delta_r$'s. The corresponding non parametric (smooth kernel) estimator of the density is given by a convolution method:
\begin{eqnarray*}
f^{np}_{ij}(t) &=& \frac{1}{n_{ij}}\sum_{k\in I_{ij}} \,K_b(t-S_k),  
\end{eqnarray*}
where $K_b(x)$ is a smoothing scaled kernel with bandwidth $b$. In our example we have chosen the Gaussian one, given by the density of the normal law of mean $0$ and variance $b^2$. In practice, many softwares provide already optimized version of non-parametric estimation of the density, with automatic choice of the bandwidth and of the kernel (here Gaussian). For example in $R$, this reduces to the function \texttt{density}, applied to the sample $\{S_k \when k \in I_{ij}\}$. 
In the symmetric case, the histogram reduces to:
\begin{eqnarray*}
f^{hst}_{\pm}\,(t_r)
&=& \frac{1}{\delta_r} \frac{\#\{k \in I_\pm \when t_r\leq S_k< t_{r+1} \}}{n_{\pm}}, 
\end{eqnarray*}
while the kernel estimator is given by: 
\begin{eqnarray*}
f^{np}_{\pm}(t) &=& \frac{1}{n_{\pm}} \sum_{k\in I_\pm} \, K_b(t-S_k).  
\end{eqnarray*}

\begin{figure}[h!]
\centering
		\includegraphics[width=1.00\textwidth]{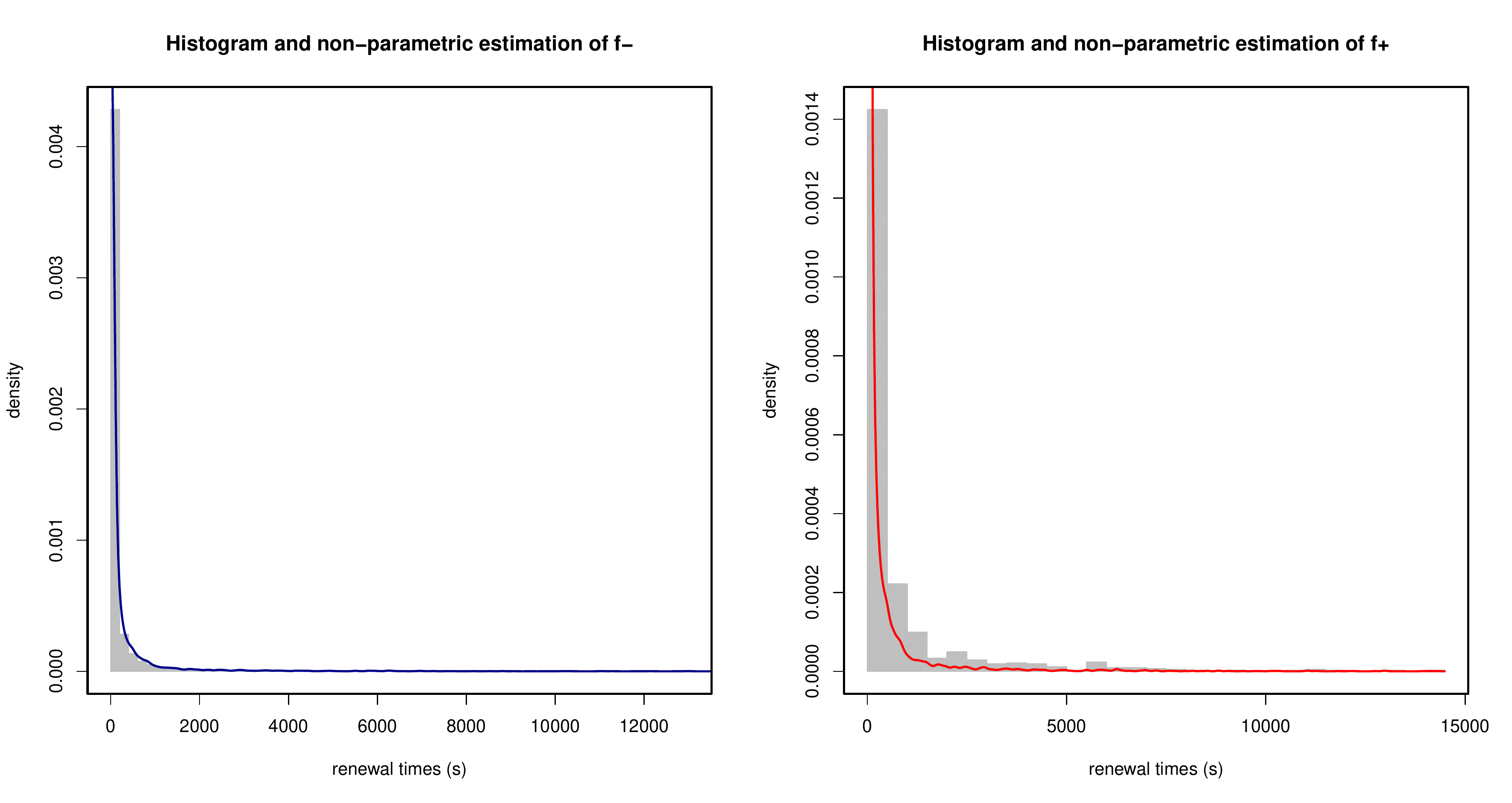}
		\caption{\small{Nonparametric estimation of  the densities $f_+$ and $f_-$ }}
		\label{densityNonParametric}
\end{figure}

Figure \ref{densityNonParametric} shows the result obtained for the kernel estimation of $f_\pm(t)$, compared to the corresponding histogram. The non parametric estimation confirms the decreasing form of both of the densities, whose interpetation is that most of the jump of the stock price takes place in a short period  of time (less than second), even though some renewal times can go to hours.\\

\begin{figure}[h!]
	\centering
		\includegraphics[width=1.00\textwidth]{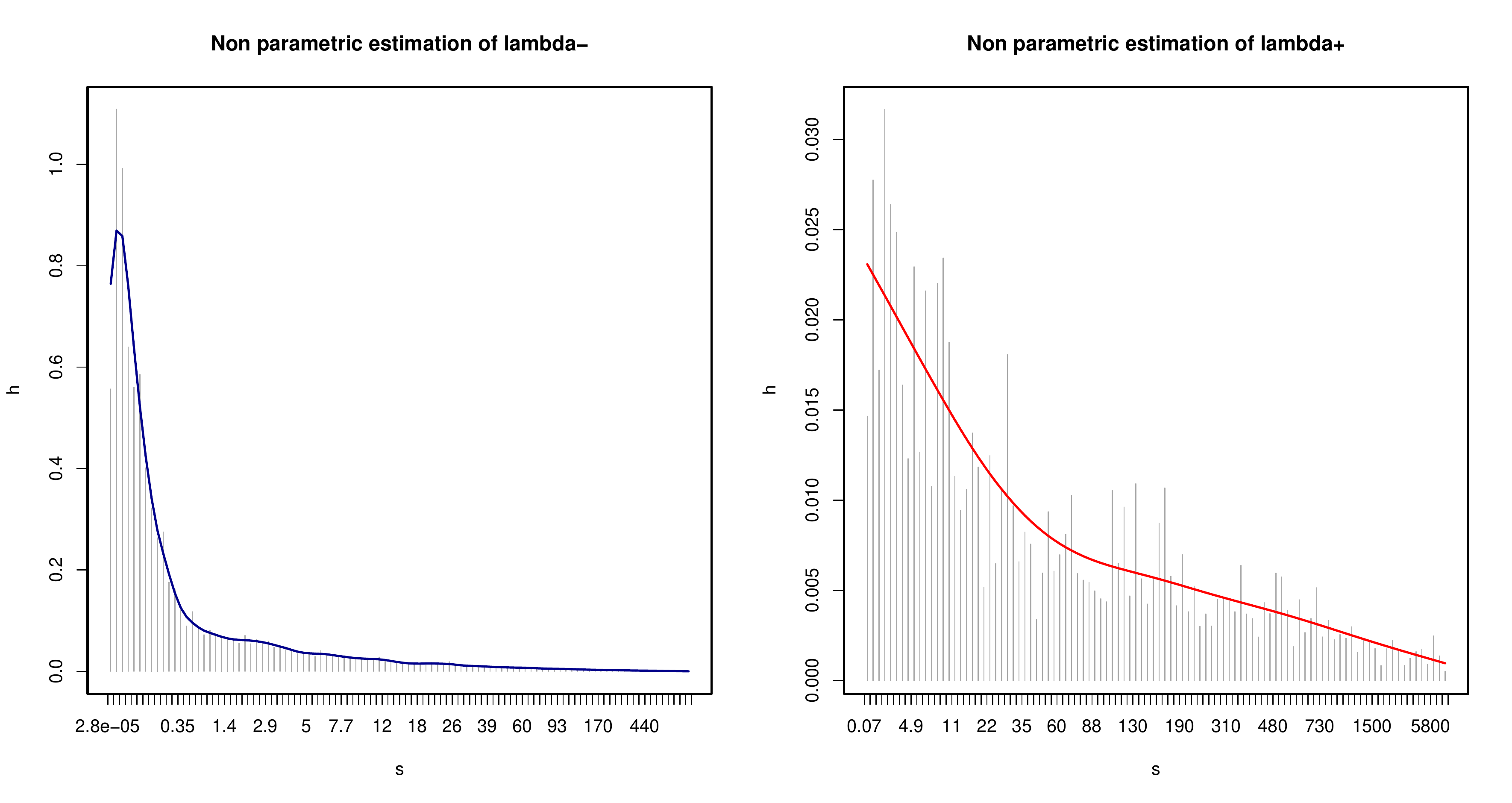}
		\caption{\small{Nonparametric estimation of the jump intensities $\lambda_+$ and $\lambda_-$}}
		\label{fig:lambdaNonParametric}
\end{figure}

We use similar technique  to estimate the marked hazard function  $h_{ij}$ defined in \reff{hazard}.  By writing 
\begin{eqnarray*}
h_{ij}(t) 
&=& \lim_{\delta \rightarrow 0}\frac{1}{\delta}
\frac{\mathbb{P}\left[t\leq S_k< t+\delta \et J_{k}=j \when J_{k-1}=i\right]} {\mathbb{P}\left[S_k\geq s \when J_{k-1}=i\right]}, 
\end{eqnarray*}
we can define the empirical histogram of $h_{ij}$ as 
\begin{eqnarray*}
h^{hst}_{ij}(t_r) 
&=& \frac{1}{\delta_r}\frac{\#\{k \in I_{ij} \when t_r\leq S_k <t_r+\delta\}}{\#\{k \in I_i \when S_k \geq t_r\}},
\end{eqnarray*} 
and the associated smooth kernel estimator by
\begin{eqnarray*}
h^{np}_{ij}(t) &=& \sum_{k\in I_{ij}} \,K_b(t-S_k)\,\frac{1}{\#\{k \in I_i \when S_k \geq t \}}. 
\end{eqnarray*} 
Notice that  this nonparametric  estimator is factorized as
\begin{eqnarray*}
h^{np}_{ij}(t) &=& \left(\frac{n_{ij}}{n_i}\right) \left(\frac{n_{i}}{n_{ij}} h^{np}_{ij}(t)\right) \\
\end{eqnarray*}
where $\frac{n_{ij}}{n_{i}}$ is the estimator of $q_{ij}$ and $\frac{n_i}{n_{ij}} h^{np}_{ij}(t)$ is the kernel estimator of the jump intensity 
$\lambda_{ij}(t)$ $=$ $\frac{f_{ij}(t)}{1-H_i(t)}$. 
Thus, we can either estimate $\lambda_{ij}$ and multiply by the estimator of $q_{ij}$ to obtain $h_{ij}$ or vice versa, obtaining the same estimators. 
In the symmetric case, we have:
\begin{eqnarray*}
h^{hst}_{\pm}(t_r) 
&=& \frac{1}{\delta_r}\frac{\#\{k \in I_\pm \when t_r\leq S_k <t_r+\delta \}}{\#\{k \when S_k \geq t_r\}},
\end{eqnarray*}
while
\begin{eqnarray*}
h^{np}_{\pm}(t) &=& \sum_{k\in I_{\pm}} \,K_b(t-S_k)\,\frac{1}{\#\{k \when S_k \geq t \}}. 
\end{eqnarray*} 
Figure \ref{fig:lambdaNonParametric} shows the result obtained for $\lambda_{\pm}$, compared to the corresponding histogram. The interpretation is the following: immediately after a jump the price is in an unstable condition, which will probably leads it to jump again soon. If it does not happen, the price gains in stability with time, and the probability of a jump becomes smaller and smaller.  Moreover,  due to mean-reversion of price returns, the intensity of  consecutive jumps in the opposite direction is larger than in the same direction, which explains  the higher value of $\lambda_-$ compared to $\lambda_+$.

\vspace{1mm}


\subsection{Price simulation}

\begin{figure}[h!] 
\centering
\includegraphics[width=9cm,height=9cm]{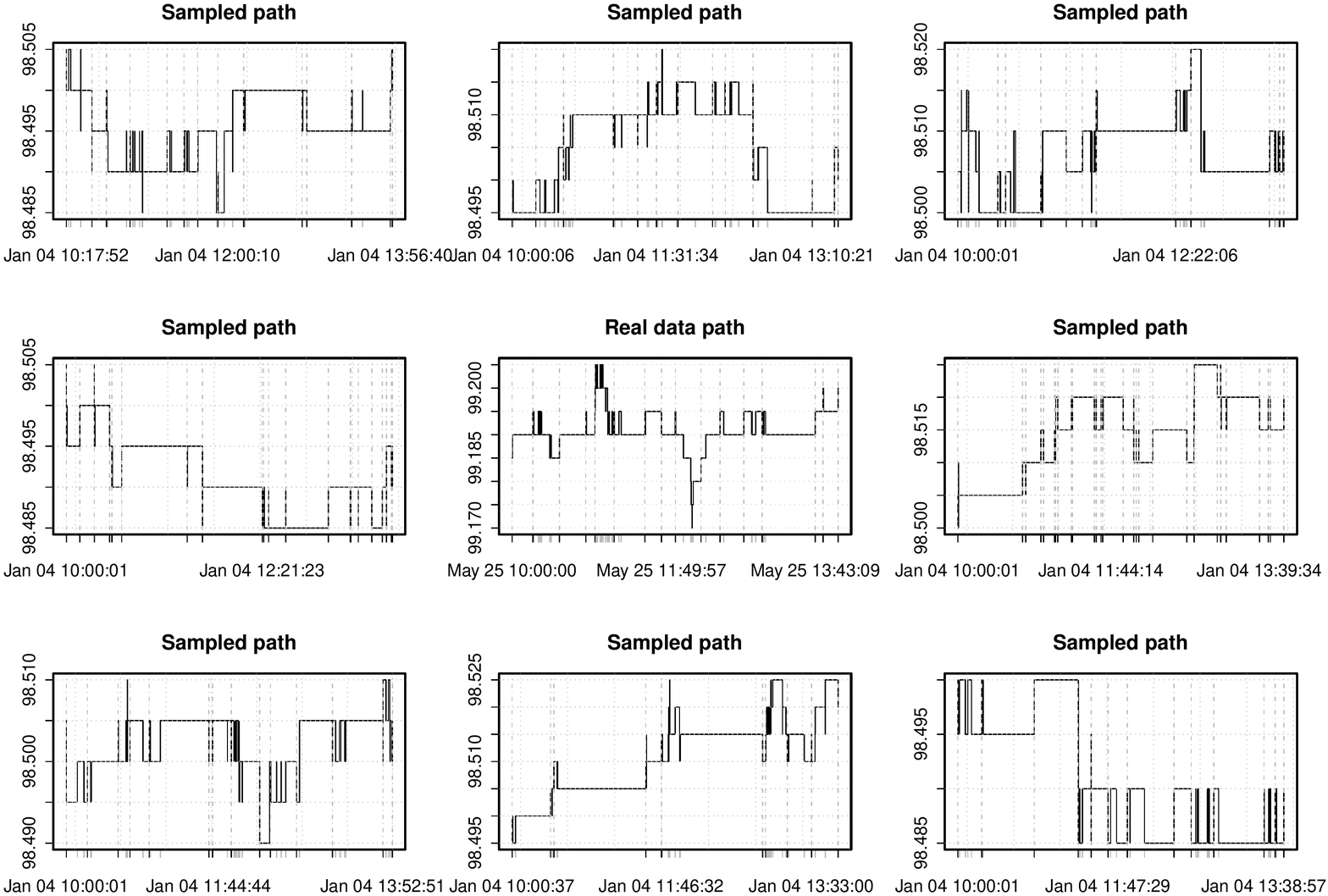} 
\caption{\small{Simulation of $9$ trajectories of the Euribor with estimated parameters on empirical data}}
\label{proratamatching} 
\end{figure}

The  simulation of the price process \reff{price} with a Markov renewal model  is quite simple, much easier than any other point process modeling microstructure noise. This would allow the user, even when $N_t$ is better fit by a more complex point process, to have a quick  proxy for its price simulation. Choose a starting price $P_0$ at initial time  $T_0$. 
\begin{description} 
	\item[Initialization step:]
	\begin{itemize}
	\item Set $\hat P_0=P_0$
	\item draw $J_0$ from initial (e.g. stationary) law $\pi$
	\end{itemize}
	\item[Inductive step:] $k-1$ $\rightarrow$ $k$ (next price and next timestamp)	
\begin{itemize}
	\item Draw $J_k$ according to the probability transition, and set $\hat P_k$ $=$ $\hat P_{k-1}$ $+$ $J_k$. 
	\item  Draw $S_k$ $\leadsto$ $F_{J_{k-1} J_k}$, and set $T_k$ $=$ $T_{k-1}+S_k$. 
\end{itemize}
\end{description}
Once $(T_k,\hat P_k)_{k\in\mathbb{N}}$ is known, the price process is given by the piecewise constant process:
\beqs
P_t &=& \hat P_k, \;\;\; T_k \leq t < T_{k+1}. 
\enqs
We show in Figure \ref{proratamatching} some simulated trajectories of the price process based on the estimated parameters of the Euribor3m.

\subsection{Semi-Markov property}

Let us  define the pure jump process, 
\begin{equation}
I_t = J_{N_t}
\label{eq:semiMarkov}
\end{equation}
which represents the last price increment. Then, $I$ is a {\it semi-Markov} process in the sense that the pair  $(I_t,S_t)$ is a Markov process, where  
\beqs
S_t &=& t -T_n, \;\; T_n \leq t < T_{n+1},
\enqs
is the time spent from the last jump price. 
Moreover, the price process $(P_t)$  is embedded in a Markov process with three observable state variables:  $(P_t,I_t,S_t)$ is a Markov process with infinitesimal generator:
\beqs
\Lc \varphi(p,i,s) &=&  \Ds{\varphi}  \; + \;     \sum_{j\in E}   h_{ij}(s)   [ \varphi(p+j,j,0) - \varphi(p,i,s) ],
\enqs
which is written in the symmetric case \reff{sym} and \reff{symF} as
\beqs
\Lc \varphi(p,i,s) &=&  \Ds{\varphi}  \; + \;    \lambda_{+}(s) \Big( \frac{1+\alpha}{2} \Big) 
 \sum_{j=1}^m p_j  [ \varphi(p+{\rm sign}(i)j,{\rm sign}(i)j,0) - \varphi(p,i,s) ] \\
 & & \;\;\;\;  + \;  \lambda_{-}(s) \Big( \frac{1-\alpha}{2} \Big)  \sum_{j=1}^m p_j  [ \varphi(p-{\rm sign}(i)j,-{\rm sign}(i)j,0) - \varphi(p,i,s) ], 
\enqs
where $\lambda_\pm$  is the jump intensity defined in \reff{jump+-}.

\subsection{Comparison with respect to Hawkes processes}

In a recent work \cite{bacetal10} (see also \cite{fautud12}), 
the authors consider a tick-by-tick model for asset price by means of cross-exciting Hawkes processes.  
More precisely, the stock price  jumps  by one tick according to the dynamics:
\beq \label{hawkes}
P_t &=& N_t^+ - N_t^-,
\enq
where $N^{\pm}$ is a point process corresponding to the number of upward and downward jumps, with coupling stochastic intensities: 
\beq \label{hawkesintensity}
\lambda_t^{\pm} &=& \lambda_\infty + \int_{-\infty}^t \varphi(t-u) dN_u^{\mp}, 
\enq
where $\lambda_\infty$ $>$ $0$ is an exogenous constant intensity, and $\varphi$ is a positive decay kernel.  This model, which takes into account  in the intensity the whole past of the price process, provides a better good-fit of intraday data than the MRP approach, 
and can be easily  extended to the multivariate asset price case. However, it is limited to unitary jump size, and 
presents some drawbacks especially in view of applications to trading optimization problem.  The price process in  \reff{hawkes}-\reff{hawkesintensity} is embedded into a  Markovian framework only for the case of exponential decay kernel, i.e. for 
\beqs
\varphi(t) &=& \gamma e^{-\beta t} 1_{\R_+}(t),
\enqs
with $0$ $<$ $\gamma$ $<$ $\beta$. In this case, the Markov price system consists of $P$ together with the stochastic intensities 
$(\lambda^+,\lambda^-)$. But in contrast with the MRP approach, the additional state variables $(\lambda^+,\lambda^-)$ are not directly observable, and have to be computed from their dynamics \reff{hawkesintensity}, which requires a ``precise" estimation of  the three parameters $\lambda_\infty$, 
$\gamma$, and $\beta$. In that sense, the MRP approach  is more robust than the Hawkes approach when dealing with Markov optimization problem. 
Notice also that in the MRP approach, we can deal not only with parametric forms of the renewal distributions (which involves only two parameters for the usual Gamma and Weibull laws), but also with non parametric form of the renewal distributions, and so of the jump intensities. Simulation and estimation in MRP model are simple since they are essentially based on (conditional) i.i.d. sequence of random variables, with the counterpart  that MRP model can not reproduce the correlation between  inter-arrival jump times as in Hawkes model.

We finally  mention that one can use a combination  of Hawkes process and semi Markov model by considering a counting process $N_t$ associated to the jump times $(T_n)$, independent of the price return  $(J_n)$, and with stochastic intensity 
\beq \label{Hawkes}
\lambda_t &=& \lambda_\infty + \gamma \int_{0}^t e^{-\beta(t-u)} dN_u,
\enq
with $\lambda_\infty$ $>$ $0$, and $0<\gamma<\beta$.  In this case, the pair $(I_t,\lambda_t)$ is a Markov process, and 
$(P_t,I_t,\lambda_t)$ is a Markov process with infinitesimal generator:
\beqs
\Lc \varphi(p,i,\ell) &=& \beta(\mu-\ell) \Dl{\varphi}  \;  + \;   \ell  \sum_{j\in E} q_{ij}  [ \varphi(p+j,j,\ell + \gamma) - \varphi(p,i,\ell) ],
\enqs
which is written in the symmetric case \reff{sym} as:
\beqs
\Lc \varphi(p,i,\ell) &=& \beta(\mu-\ell) \Dl{\varphi}  \\
& & \;\;\; + \;   \ell  \Big( \frac{1+\alpha}{2} \Big) \sum_{j=1}^m p_j  [ \varphi(p+{\rm sign}(i)j,{\rm sign}(i)j,\ell + \gamma) - \varphi(p,i,\ell) ] \\
& &  \;\;\; + \;   \ell  \Big( \frac{1- \alpha}{2} \Big) \sum_{j=1}^m p_j  [ \varphi(p-{\rm sign}(i)j,-{\rm sign}(i)j,\ell + \gamma) - \varphi(p,i,\ell) ]. 
\enqs

\section{Scaling limit}\label{section:diffusive}

\setcounter{equation}{0} \setcounter{Assumption}{0}
\setcounter{Theorem}{0} \setcounter{Proposition}{0}
\setcounter{Corollary}{0} \setcounter{Lemma}{0}
\setcounter{Definition}{0} \setcounter{Remark}{0}

We now study the large scale limit of the price process constructed from our Markov renewal model.  We consider the symmetric case 
\reff{sym} and \reff{symF}, and denote by 
\beqs
F(t) &=& \frac{1+\alpha}{2} F_+(t) +  \frac{1- \alpha}{2} F_-(t),
\enqs
the distribution function of the sojourn time $S_n$.  We assume that the mean sojourn time is finite: 
\beqs
\bar \mu &=& \int_0^\infty t \; dF(t) \; < \; \infty, \;\;\; \mbox{ and we set } \;  \bar\lambda = \frac{1}{\bar\mu}. 
\enqs
By classical regenerative arguments, it is known  (see \cite{glyhas04}) that  the Markov renewal process  obeys a strong law of large numbers, which means  the long run stability of price process:
 \beqs
 \frac{P_t}{t} & \longrightarrow & c, \;\;\; a.s. 
 \enqs
 when $t$ goes to infinity, with a limiting constant $c$ given by:
 \beqs
 c &=& \frac{1}{\bar\mu}  \sum_{i,j \in E}  \pi_i\, q_{ij}\, j, 
 \enqs
 where $E$ $=$ $\{1,-1,\ldots,m,-m\}$, $\bar\mu_{ij}$ $=$ $\int_0^\infty t dF_{ij}(t)$, $Q$ $=$ $(q_{ij})$ is the transition matrix \reff{Qparti} of the embedded Markov chain $(J_n)_n$, and  $\pi$ $=$ $(\pi_i)$ is the invariant distribution of $(J_n)_n$.   
In the symmetric case \reff{sym} and \reff{symF},  we have $q_{ij}$ $=$ $p_{|i|}(1+{\rm sign}(j)\alpha)/2$, $\pi_i$ $=$ $p_{|i|}/2$, and so $c$ $=$ $0$.
We next define the normalized price process:
\beqs
P^{(T)}_t &=& \frac{P_{tT}}{\sqrt{T}}, \;\;\; t \in [0,1],
\enqs
and address the macroscopic limit of $P^{(T)}$ at large scale limit $T$ $\rightarrow$ $\infty$.  From the functional central limit theorem for Markov renewal  process,  we  obtain the large scale  diffusive behavior of price process.

\begin{Proposition} \label{propdiff}
\beqs
\lim_{T\rightarrow \infty} P^{(T)}  & \stackrel{ (d) }{=} & \sigma_{_\infty} W, 
\enqs
where $W$ $=$ $(W_t)_{t\in [0,1]}$ is a standard Brownian motion, and $\sigma_{_{\infty}}$, the macroscopic variance,  is 
explicitly given by
\beq \label{sigmamacro}
\sigma_{_\infty}^2 &=& \bar\lambda \Big[ {\rm Var}(\xi_n) + \big(\E[\xi_n]\big)^2 \; \frac{1+\alpha}{1-\alpha} \Big].   
\enq
\end{Proposition}
{\bf Proof.}
From  \cite{glyhas04}, we know that  a functional central limit theorem holds for Markov renewal process so that 
\beqs
P^{(T)} &  \stackrel{ (d) }{\rightarrow}  & \sigma_{_\infty} W
\enqs
when $T$ goes to infinity (here $\stackrel{ (d) }{\rightarrow}$ means convergence in distribution), where $\sigma_{_\infty}$ is given by 
$\sigma_{_\infty}^2$ $=$  $\bar\lambda \tilde\sigma_{_\infty}^2$ with  $\bar\lambda$ $=$ $1/\bar\mu$, and 
\beqs
\tilde\sigma_{_\infty}^2 &=&  \sum_{i,j\in E} \pi_i q_{ij} H_{ij}^2,
\enqs
with
\beqs
H_{ij} &=& j + g_j - g_i \\
g = (g_i)_i &=& (I_{2m} - Q + \Pi)^{-1} b \\
b = (b_i), \; b_i &=& \sum_{j\in E}  Q_{ij} j,
\enqs
and $\Pi$ is the matrix defined by $\Pi_{ij}$ $=$ $\pi_j$.  In the symmetric case \reff{sym}, a straightforward calculation shows that  
\beqs
b_i &=& {\rm sign}(i) \alpha \sum_{j=1}^m j p_j \; = \; {\rm sign}(i) \alpha \E[\xi_n],
\enqs
and then
\beqs
g_i &=& {\rm sign}(i) \frac{\alpha}{1-\alpha} \E[\xi_n],  \;\;\; i \in E= \{1,-1,\ldots,m,-m\}.  
\enqs
Thus, 
\beqs
H_{ij} &=&  \left\{ 
		\begin{array}{cc}
		j & \mbox{ if } ij > 0 \\
		j  +   \frac{2\alpha}{1-\alpha} \E[\xi_n] & \mbox{ if }  j>0, i<0 \\
		j  -   \frac{2\alpha}{1-\alpha} \E[\xi_n] & \mbox{ if }  j<0, i>0. 
		\end{array}
		\right. 
\enqs
Therefore, a direct calculation yields
\beqs
\tilde\sigma_{_\infty}^2 &=& \sum_{j=1}^m j^2 p_j  +  \frac{2\alpha}{1-\alpha} \big( \E[\xi_n]  \big)^2 \\
&=&  {\rm Var}(\xi_n) + \big(\E[\xi_n]\big)^2 \; \frac{1+\alpha}{1-\alpha}. 
\enqs
\ep

 \section{Mean Signature plot}\label{section:signature}

\setcounter{equation}{0} \setcounter{Assumption}{0}
\setcounter{Theorem}{0} \setcounter{Proposition}{0}
\setcounter{Corollary}{0} \setcounter{Lemma}{0}
\setcounter{Definition}{0} \setcounter{Remark}{0}

In this section, we aim to provide through our Markov renewal model a quantitative justification of the signature plot effect, described in the introduction.   
We consider the symmetric and stationary case, i.e.:

\vspace{2mm}

\noindent {\bf (H)} 
\begin{itemize}
\item[(i)]  The price return $(J_n)_n$  is given by  \reff{Jn} with a probability transition matrix $\hat Q$ for $(\hat J_n)_n$ $=$ 
$({\rm sign}(J_n))_n$ in the form: 
\beqs
\hat Q &=& \left( \begin{array}{cc}
			   \frac{1+\alpha}{2} & \frac{1-\alpha}{2} \\
			   \frac{1- \alpha}{2} & \frac{1 + \alpha}{2} 
			   \end{array}
		  \right) 
\enqs
for some $\alpha$ $\in$ $[-1,1)$ . 
\item[(ii)] $(N_t)$ is a delayed renewal process:  for $n$ $\geq$ $1$, $S_n$ has distribution function $F$ (independent of $i,j$),  with finite mean 
$\bar\mu$ $=$ $\int_0^\infty t \; dF(t)$ $=:$ $1/\bar\lambda$  $<$ $\infty$, and finite second moment, and $S_1$ has a distribution with 
density $(1-F(t))/\bar\mu$. 
\end{itemize}

It is known that the process $N$ is stationary under {\bf (H)}(ii) (see \cite{daljon02}), and so the price process is also stationary under the  
stationary probability $\P_\pi$, i.e. the distribution of $P_{t+\tau}-P_t$  does not depend on $t$ but only on increment time $\tau$. 
In this case, the empirical mean signature plot is  written as:
\beq
\bar V(\tau) &:= &  \frac{1}{T} \sum_{i\tau\leq T} \mathbb{E}_\pi \big[\big(P_{i\tau}-P_{(i-1)\tau}\big)^2\big] \nonumber \\
&=& \frac{1}{\tau} \E_\pi \big[ \big( P_\tau - P_0\big)^2 \big]  \label{meansignature}
\enq
Notice that  if $P_t=\sigma W_t$, where $W_t$ is a Brownian motion, then $\bar V$ is a flat function: $\bar V(\tau)=\sigma^2$, while it is well known that on real data $\bar V$ is a decreasing function on $\tau$ with finite limit when  $\tau\rightarrow\infty$. 
This is mainly due to the anticorrelation of returns: on a short time-step the signature plot captures fluctuations due to returns that, on a longer time-steps, mutually cancel. We obtain the closed-form expression  for the mean signature plot, and give some qualitative properties about the 
impact of price returns autocorrelation. The following results are proved in Appendix.

\begin{Proposition}  \label{propsignature}
Under {\bf (H)}, we have: 
\beqs
\bar V(\tau) &=&  \sigma_\infty^2  +  \Big(  \frac{-2\alpha (\E [\xi_n])^2}{1-\alpha}\Big)  \frac{1- G_\alpha(\tau)}{(1-\alpha)\tau},
\enqs
where  $\sigma_{_\infty}^2$ is the macroscopic variance given in \reff{sigmamacro},  and  $G_\alpha(t)$ $:=$ $\E[\alpha^{N_t}]$ is given  via its Laplace-Stieltjes transform:
\beq \label{laplaceG}
\widehat G_\alpha(s) \; = \;  1 -  \bar\lambda(1-\alpha) \frac{1-\widehat F(s)}{s(1-\alpha\widehat F(s))},  \;\;\; \alpha \neq 0,
\enq
$\widehat F(s)$ $:=$ $\int_{0^-}^\infty e^{-st} dF(t)$.  Alternatively,  $G_\alpha$  is given directly by the integral form: 
\beq \label{integG}
G_\alpha(t) &=& 1 - \bar\lambda\Big(\frac{1-\alpha}{\alpha}\Big)\Big( t - (1-\alpha) \int_0^t Q_\alpha^0(u) du  \Big),
\enq
where $Q_\alpha^0(t)$ $=$ $\Sum_{n=0}^\infty \alpha^n F^{*(n)}(t)$,  and $F^{*(n)}$ is the $n$-fold convolution of the distribution function $F$, i.e. 
$F^{*(n)}(t)$ $=$ $\int_{0}^t  F^{*(n-1)}(t-u)dF(u)$, $F^{*(0)}$ $=$ $1$. 
\end{Proposition}

\vspace{1mm}

\begin{Corollary}
Under {\bf (H)}, we obtain the asymptotic   behavior of the mean signature plot: 
\beqs
\bar V(\infty) \; := \;  \lim_{\tau\rightarrow\infty} \bar V(\tau) &=& \sigma^2_{_\infty}, \\
\bar V(0^+) \; := \;  \lim_{\tau\downarrow 0^+} \bar V(\tau) &=&  \bar\lambda \E[\xi_n^2].
\enqs
Moreover, 
\beqs
\alpha \big( \bar V(0^+) -  \bar V(\infty) \big) & \leq & 0. 
\enqs
\end{Corollary}

\vspace{2mm}

\begin{Remark}
{\rm    In the case of renewal process where $F$ is the distribution function of the Gamma law  with shape $\beta$ and scale $\theta$,  it is known that  $F^{*(n)}$ is the distribution function a the Gamma law with shape $n\beta$ and scale $\theta$, and so: 
\beqs
F^{*(n)}(t) &=& \frac{\Gamma_{t/\theta}(n\beta)}{\Gamma(n\beta)},
\enqs
where $\Gamma$ is the Gamma function, and $\Gamma_t$ is the  lower incomplete Gamma functions defined in \reff{defgamma}.  
Plugging into \reff{integG}, we obtain an explicit integral expression of the mean signature plot, which is computed numerically by avoiding the inversion of the Laplace transform \reff{laplaceG}. Notice that in the special case of Poisson process for $N$, i.e.  $F$ is the exponential distribution of rate 
$\bar\lambda$,  the function $G_\alpha$ is explicitly given by : 
$G_\alpha(t)$ $=$ $e^{-\bar\lambda(1-\alpha)t}$. 
}
\end{Remark}

\begin{figure}[h!] 
\centering
{
\includegraphics[width=6.7cm,height=6.3cm]{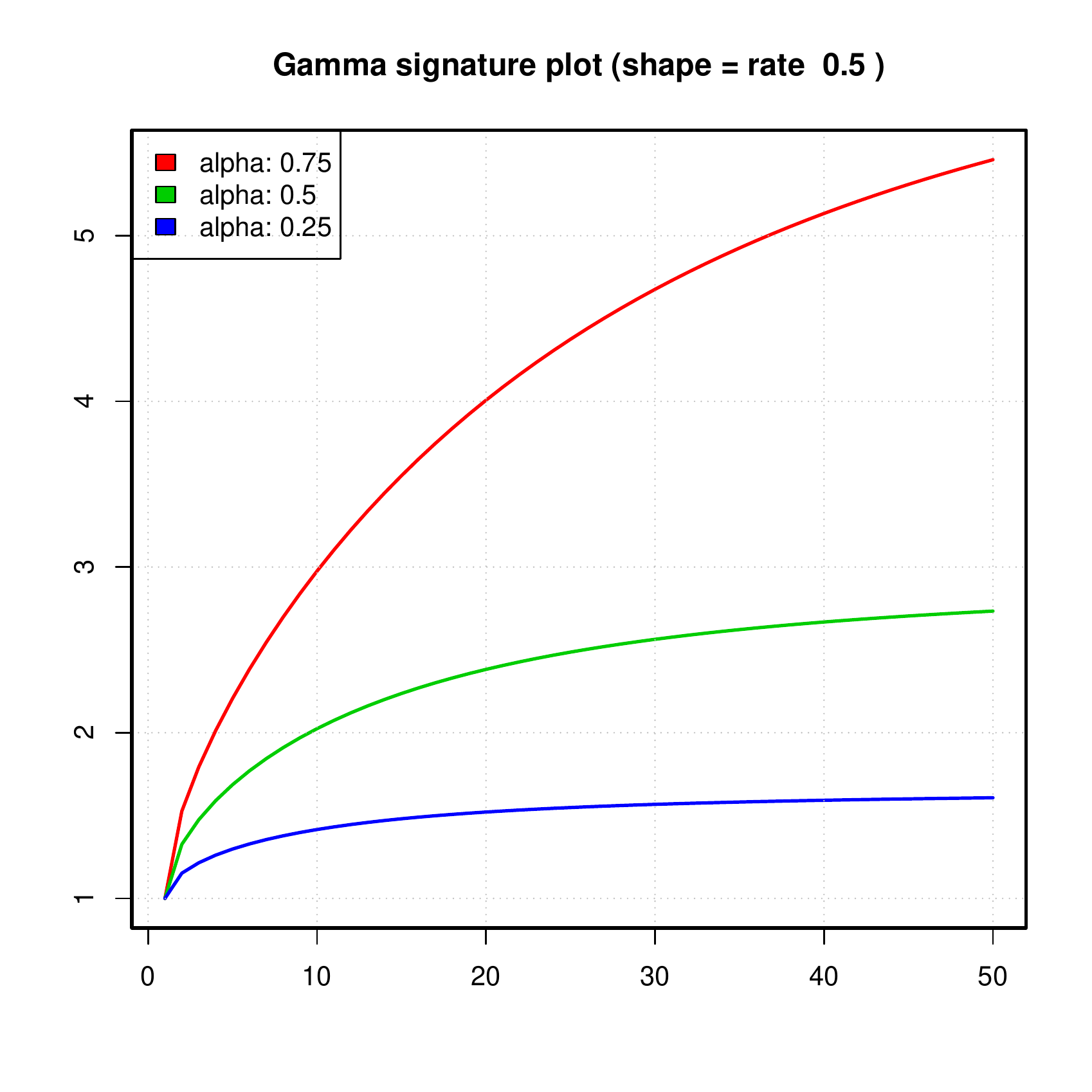}  
}
{
\includegraphics[width=6.7cm,height=6.3cm]{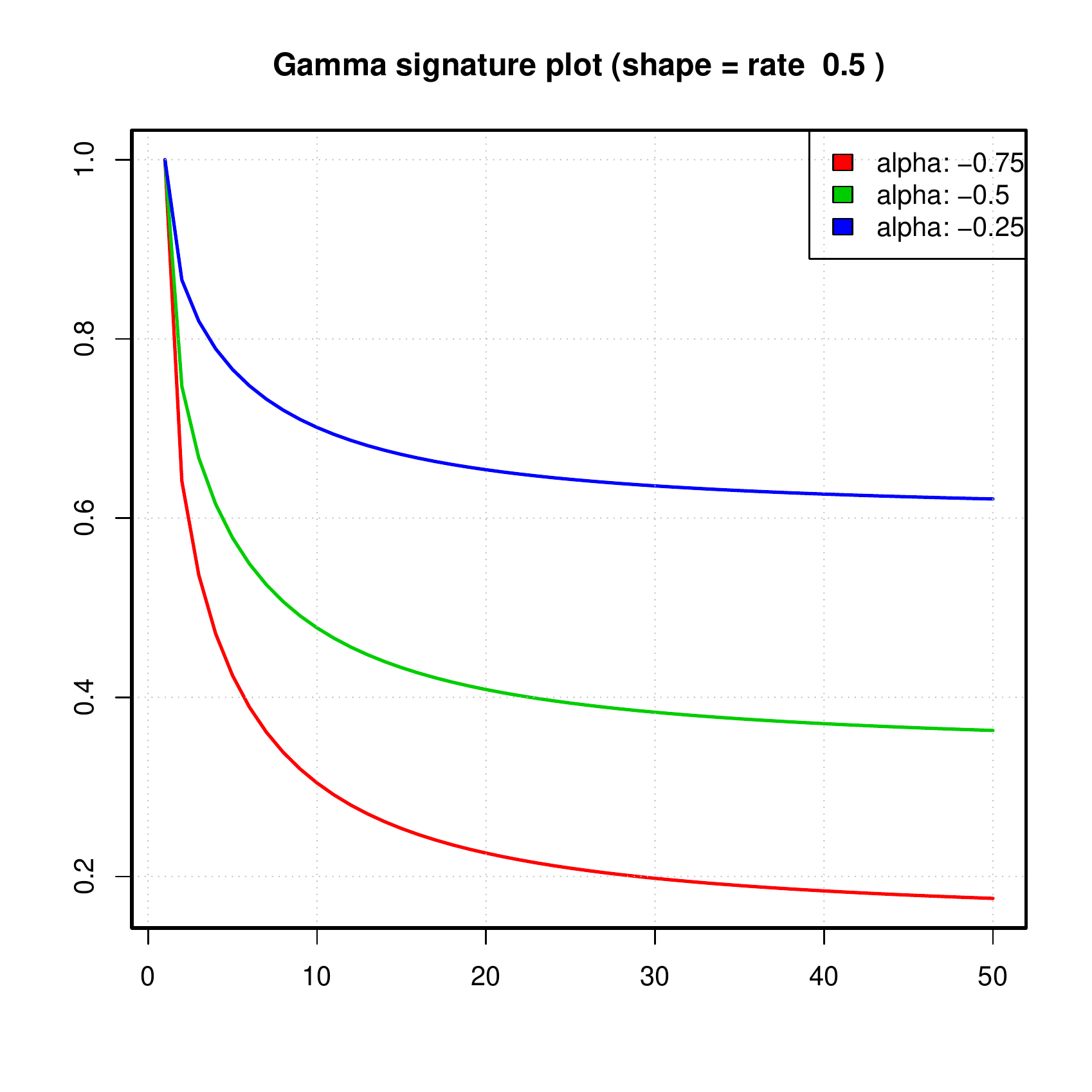} 
}
\caption{\small{$\tau$ $\rightarrow$ $\bar V(\tau)$ when varying price return autocorrelation $\alpha$. Left: $\alpha$ $>$ $0$. Right: $\alpha$ $<$ $0$}}
\label{signaturealpha}
\end{figure}

\begin{figure}[h!] 
\centering
\includegraphics[width=7.5cm,height=7.5cm]{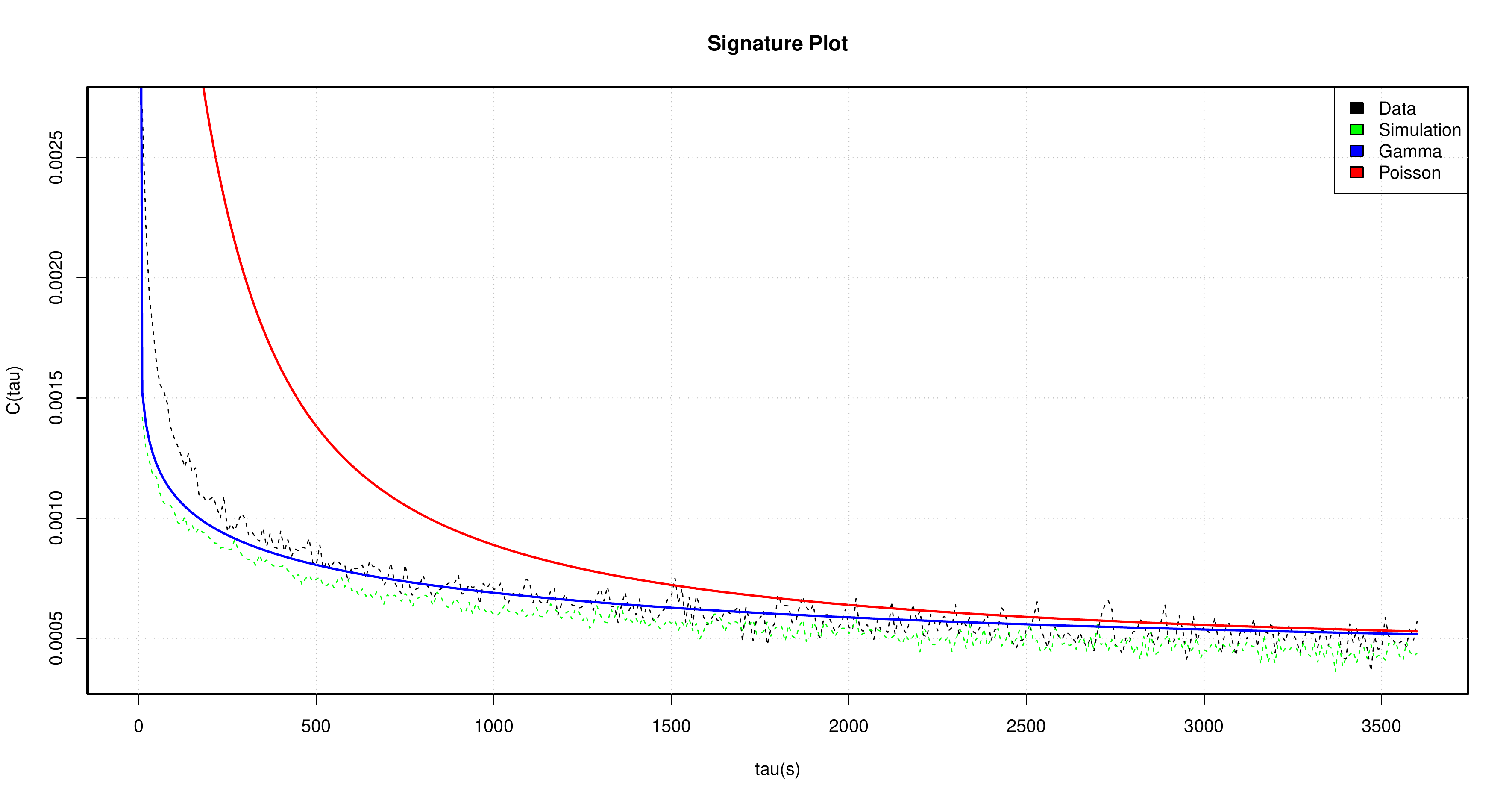} 
\caption{\small{Comparison of signature plot: empirical data, simulated and computed}}
\label{signaturecomp}
\end{figure}

\begin{Remark}
{\rm  The term $\sigma^2_{_\infty}$ equal to the limit of the mean signature plot when time step observation $\tau$ goes to infinity, corresponds to the macroscopic variance, and 
$\bar V(0^+)$ $=$  $\bar\lambda \E[\xi_n^2]$ is the microstructural variance. Notice that while $\sigma^2_{_\infty}$ increases with  the price returns autocorrelation $\alpha$,  the limiting term $\bar V(0^+)$ does not depend on $\alpha$, and  the mean signature plot is flat if and only if  price returns are  independent, i.e. $\alpha$ $=$ $0$.  In the case of mean-reversion ($\alpha$ $<$ $0$),  
$\bar V(0^+)$ $>$ $\sigma^2_{_\infty}$, while in the trend case ($\alpha$ $>$ $0$), we have: $\bar V(0^+)$ $<$ $\sigma^2_{_\infty}$. We display in Figures  \ref{signaturealpha}  plot example  of   the mean signature plot function for a Gamma distribution when varying $\alpha$. We also compare in Figure \ref{signaturecomp} the signature plot obtained from empirical data on the Euribor, the signature plot simulated in our model with estimated parameters, and the mean signature for a Gamma distribution with the estimated parameters. This example shows how the signature plot decreasing form is a consequence (rather coarse) of the microstructure noise, and that the shape parameters of the gamma law, which is responsible for the volatility cluster, is able to reproduce the convexity of the signature plot shape. 
}
\end{Remark}

\section{Conclusion and extensions}

\setcounter{equation}{0} \setcounter{Assumption}{0}
\setcounter{Theorem}{0} \setcounter{Proposition}{0}
\setcounter{Corollary}{0} \setcounter{Lemma}{0}
\setcounter{Definition}{0} \setcounter{Remark}{0}

In this paper we used a Markov renewal process $(T_n,J_n)_n$ to describe the tick-by-tick evolution of the stock price and reproduce two important stylized facts, as the diffusive behavior and the decreasing shape of the mean signature plot towards the diffusive variance. Ha\-ving in mind a direct application purpose, we decided to sacrifice the autocorrelation of inter-arrival times in order to have a fast  and simple  non-parametric estimation, perfect simulation and the 
suitable setup for a market making application, presented in a companion paper \cite{fodpha13b}. 
Aware of the model limits, and with an eye to statistical arbitrage, the next step is  to extend the structure of the counting process $N_t$, for example  to  Hawkes process,  to have a better fit, and the structure of the Markov chain $(J_n)_n$ to a longer memory binary processes, able to recognize patterns. Another direction of study will be the extension to the multivariate asset price case.

\appendix

\section{Appendix: Mean signature plot}
 
 \setcounter{equation}{0} \setcounter{Assumption}{0}
\setcounter{Theorem}{0} \setcounter{Proposition}{0}
\setcounter{Corollary}{0} \setcounter{Lemma}{0}
\setcounter{Definition}{0} \setcounter{Remark}{0}

The price  process $P_t$ is  given  by 
\beq \label{prixP}
P_t &=& P_0 + \sum_{k=1}^{N_t} J_k \; = \; \sum_{n=0}^\infty L_n 1_{N_t=n}, 
\enq
where
\beqs
L_n &=& \sum_{k=1}^n J_k, \;\;\;n \geq 1, \;\; L_0 \; = \; 0.
\enqs

\begin{Lemma}  \label{lemLn}
Under {\bf (H)}, we have
\beqs
\E_{\pi} \big[ L_n^2 \big] &=& n \frac{\sigma_{_\infty}^2}{\bar\lambda}  - \frac{2\alpha}{1-\alpha} \frac{1-\alpha^n}{1-\alpha}  \big(\E[\xi_n]\big)^2, 
\enqs
where $\sigma_{_\infty}$ is defined in \reff{sigmamacro}. 
\end{Lemma}
{\bf Proof.} By writing that $L_n$ $=$ $L_{n-1}$ $+$ $\xi_n \hat J_n$,  we have  for $n$ $\geq$ $1$
\beq
\E_\pi \big[ L_n^2 \big] &=& \E_\pi\big[ L_{n-1}^2 + \xi_n^2 \hat J_n^2  + 2 \xi_n L_{n-1} \hat J_n \big]  \nonumber \\
&=& \E_\pi \big[ L_{n-1}^2\big] + \E_{\pi}\big[ \xi_n^2\big]  + 2 \big(\E\big[ \xi_n\big]\big)^2 \sum_{k=1}^{n-1}  \E_\pi \big[ \hat J_{k}\hat J_n\big], \nonumber \\
&=&  \E_\pi \big[ L_{n-1}^2\big] + \E_{\pi}\big[ \xi_n^2\big]  + 2 \big(\E\big[ \xi_n\big]\big)^2 \sum_{k=1}^{n-1}  \E_\pi \big[ \hat J_{k} \hat J_0\big] \label{interLn} 
\enq
where we used the fact that $\hat J_n^2$ $=$ $1$,  $(\xi_n)_n$ are i.i.d, and independent of $(\hat J_n)_n$ in the second equality, and the stationarity of $(\hat J_n)_n$ in the third equality.  Now, from the Markov property of $(\hat J_k)_k$ with probability transition matrix $\hat Q$ in \reff{hatQmatrix} and \reff{sym}, we have  for any $k$ $\geq$ $1$:
\beqs
\E_\pi\big[ \hat J_k \hat J_0 \big] &=& \E_\pi \Big[ \E_\pi\big[ \hat J_k | \hat J_{k-1}\big] \hat J_0 \Big] \\
&=& \E \Big[ \Big(\frac{1+\alpha}{2}\Big) \hat J_{k-1} \hat J_0 -  \Big(\frac{1- \alpha}{2}\Big) \hat J_{k-1} \hat J_0 \Big] \\
&=& \alpha \E_\pi \big[ \hat J_{k-1}\hat J_0 \big], 
\enqs
from which we obtain by induction:
\beqs
\E_\pi\big[ \hat J_k \hat J_0 \big] &=& \alpha^k.
\enqs
Plugging into  \reff{interLn}, this gives
\beqs
\E_\pi \big[ L_n^2 \big] &=& \E_\pi \big[ L_{n-1}^2\big] + \E_{\pi}\big[ \xi_n^2\big]  +  \frac{2\alpha}{1-\alpha} (1-\alpha^{n-1}) \big(\E[\xi_n]\big)^2.
\enqs
By induction, we get the required relation for $\E_\pi \big[ L_n^2 \big]$.   
\ep

\vspace{2mm}

Consequently, we obtain the following expression of the mean signature plot: 

\begin{Proposition}
Under {\bf (H)}, we have
\beq \label{Vsign}
\bar V(\tau) &=& \sigma_{_\infty}^2   -  \frac{2\alpha}{1-\alpha} \frac{1-G_\alpha(\tau)}{(1-\alpha)\tau}  \big(\E[\xi_n]\big)^2, \;\;\; \tau >  0, 
\enq
where $G_\alpha(t)$ $:=$ $\E_\pi[\alpha^{N_t}]$ $=$ $\sum_{n=0}^\infty \alpha^n \P_\pi[N_t = n]$. 
\end{Proposition}
{\bf Proof.} From \reff{meansignature} and \reff{prixP}, we see that the mean signature plot is written as
 \beqs
\bar V(\tau) &=& \frac{1}{\tau} \sum_{n=0}^\infty \E_\pi[L_n^2] \P_\pi [N_\tau = n], 
\enqs
since the renewal process $N$ is independent of the marks $(J_n)$. Together with the expression of $\E_\pi[L_n^2]$ in Lemma \ref{lemLn},  this yields 
\beqs
\bar V(\tau) &=& \frac{\sigma_{_\infty}^2}{\bar\lambda}  \frac{\E_\pi[N_\tau]}{\tau}  -  \frac{2\alpha}{1-\alpha} \frac{1-G_\alpha(\tau)}{(1-\alpha)\tau}  \big(\E[\xi_n]\big)^2.
\enqs
Finally, since $\E_\pi[N_\tau]$ $=$ $\bar\lambda\tau$ by stationarity of  $N$, we get the required relation.  
\ep

\vspace{3mm}

We now focus on the finite variation function $G_\alpha$ that we shall compute through its Laplace-Stieltjes transform:
\beqs
\widehat G_\alpha(s) &:=& \int_{0^-}^\infty e^{-st} dG_\alpha(t), \;\;\; s \geq 0. 
\enqs
We recall  the convolution property for Laplace-Stieltjes transform
\beqs
\widehat{G*H} &=& \widehat G. \widehat H,
\enqs
where
\beqs
G*H(t) &=& \int_{0}^t G(t-s) dH(s). 
\enqs
Let us consider the function $Q_\alpha$ defined by:
\beq \label{defQ}
Q_\alpha(t)  & := &  \sum_{n=0}^\infty \alpha^n \P_\pi [N_t \geq n]. 
\enq

\begin{Lemma}
Under {\bf (H)}, we have for all $\alpha$ $\neq$ $0$,
\beq
G_\alpha &=& \Big(1  - \frac{1}{\alpha}\Big) Q_\alpha + \frac{1}{\alpha} \label{GQ} \\
\widehat Q_\alpha(s) &=& 1 + \alpha \frac{\bar\lambda}{s} \Big( \frac{1-\widehat F(s)}{1-\alpha \widehat F(s)} \Big).  \label{hatQ}
\enq
\end{Lemma}
{\bf Proof.} {\bf 1.}  For any $\alpha$ $\neq$ $0$, $t$ $\geq$ $0$, we have
\beqs
G_\alpha(t) &=& \sum_{n=0}^\infty \alpha^n \P_\pi[N_t = n] \; = \; \sum_{n=0}^\infty \alpha^n \big( \P_\pi[N_t \geq n] - \P_\pi[N_t \geq n+1] \Big) \\
&=&   \sum_{n=0}^\infty \alpha^n  \P_\pi[N_t \geq n]  -  \frac{1}{\alpha}  \Big( \sum_{n=0}^\infty \alpha^n  \P_\pi[N_t \geq n] - \P_\pi[N_t \geq 0] \Big) \\
&=& Q_\alpha(t)  -  \frac{1}{\alpha}  \Big( Q_\alpha(t) - 1 \Big),
\enqs
which proves \reff{GQ}. 

\noindent {\bf 2.}  Recall that for the delayed renewal process $N$, the first arrival time $S_1$ is distributed according to the distribution $\Delta$ with density   $\bar\lambda(1-F)$. Let us denote by $N^0$ the no-delayed renewal process, i.e. with all interarrival times $S_n^0$ $=$ $T_n^0-T_{n-1}^0$ distributed according to $F$, and by 
$Q_\alpha^0$ the function defined similarly as in \reff{defQ} with $N$ replaced by $N^0$.   Then, 
\beq
Q_\alpha(t) &=& 1 + \alpha \sum_{n=0}^\infty \alpha^n \P_\pi\big[ N_t \geq n+1 \big]  \nonumber \\
&=& 1 +  \alpha  \sum_{n=0}^\infty \alpha^n \P_\pi [ S_1 + T_n^0 \leq t ] \nonumber \\
&=&  1 +  \alpha \E_\pi \Big[  \sum_{n=0}^\infty \alpha^n \P_\pi [ T_n^0 \leq t-S_1 | S_1 \leq t ] \Big]  \nonumber \\
&=&  1 +  \alpha \E_\pi \Big[ Q_\alpha^0(t-S_1) 1_{S_1\leq t}  \Big] \nonumber \\
&=& 1 + \alpha Q_\alpha^0*\Delta(t), \label{Qalpha}
\enq
By taking the Laplace-Stieltjes  transform in the above relation, and from the convolution property, we get 
\beq \label{laplaceQ}
\widehat Q_\alpha &=& 1 + \alpha \widehat{Q_\alpha^0}\widehat \Delta. 
\enq
By same arguments  as in  \reff{Qalpha} and \reff{laplaceQ}, we get  $\widehat{Q_\alpha^0}$ $=$ $1 + \alpha \widehat{Q_\alpha^0}\widehat  F$, and so  
\beq \label{laplaceQ0alpha}
\widehat{Q_\alpha^0} &=& \frac{1}{1-\alpha\widehat F}. 
\enq
Now, from the relation  $\Delta(t)$ $=$ $\int_0^t \bar\lambda(1-F(u))du$, and by taking Laplace-Stieltjes transform we get:
\beq \label{laplacedelta}
\widehat\Delta(s) &=& \frac{\bar\lambda}{s}\big(1-\widehat F(s) \big). 
\enq
By substituting \reff{laplaceQ0alpha} and \reff{laplacedelta} into \reff{laplaceQ}, we get the required relation \reff{hatQ}. 
\ep

\vspace{3mm}

From the relations \reff{GQ}-\reff{hatQ} in the above Lemma, we immediately obtain the expression \reff{laplaceG} for  the  Laplace-Stieltjes transform $\widehat G_\alpha$. 
 Let us now derive the alternative integral expression  \reff{integG} for $G_\alpha$.

 \begin{Lemma}
 Under {\bf (H)}, we have for all $\alpha$ $\neq$ $0$: 
 \beq \label{Qinteg}
 Q_\alpha(t) &=& 1 + \bar\lambda t - \bar\lambda(1-\alpha) \int_0^t Q_\alpha^0(u) du,
 \enq
 with 
 \beqs
 Q_\alpha^0(t) &=&  \Sum_{n=0}^\infty \alpha^n F^{*(n)}(t),  
 \enqs
 and $F^{*(n)}$ is the $n$-fold convolution of the distribution function $F$.
 \end{Lemma}
 {\bf Proof.}  We rewrite the expression \reff{hatQ} Laplace-Stieltjes transform as
 \beq
 \widehat Q_\alpha(s) &=& 1 + \frac{\alpha\bar\lambda}{s}  (1-\widehat F)  \sum_{n=0}^\infty (\alpha \widehat F)^n   \nonumber \\
 &=& 1 + \frac{\alpha\bar\lambda}{s} \sum_{n=0}^\infty \alpha^n \Big( (\widehat F)^n -  (\widehat F)^{n+1} \Big). \label{interQ}
 \enq
 Let us now consider the function $G_\alpha^0$ (resp.  $Q_\alpha^0$) defined similarly as for $G_\alpha$ (resp. $Q_\alpha$) with $N$ replaced by $N^0$, the no-delayed renewal process with all interarrival times $S_n^0$ $=$ $T_n^0-T_{n-1}^0$ distributed according to 
 $F$.  Then, 
 \beq
 G_\alpha^0(t) &=&  \sum_{n=0}^\infty \alpha^n \big( \P_\pi[N_t^0 \geq n] - \P_\pi[N_t^0 \geq n+1] \Big) \nonumber  \\
 &=& \sum_{n=0}^\infty \alpha^n \Big( \P_\pi[T_n^0 \leq t] - \P_\pi[T_{n+1}^0 \leq t] \Big) \nonumber \\
 &=&  \sum_{n=0}^\infty \alpha^n \Big( F^{*(n)}-\ F^{*(n+1)})(t).  \nonumber
 \enq
Therefore,  the Laplace-Stieltjes transform of $G_\alpha^0$ is written also as
 \beqs
 \widehat{G_\alpha^0} &=& \sum_{n=0}^\infty \alpha^n \Big( (\widehat F)^n -  (\widehat F)^{n+1} \Big).
 \enqs
 By defining the function $I_\alpha^0(t)$ $:=$ $\int_0^t G_\alpha^0(u) du$, we then see from \reff{interQ} that 
 \beqs
  \widehat Q_\alpha(s) &=& 1 +  \alpha\bar\lambda \widehat{I_\alpha^0},
 \enqs
 and thus
 \beq \label{interQ0}
 Q_\alpha(t) &=& 1 +  \alpha\bar\lambda  \int_0^t G_\alpha^0(u) du. 
 \enq
 Finally, by same arguments as in \reff{GQ}, we have 
 \beqs
 G_\alpha^0 &=& \Big(1  - \frac{1}{\alpha}\Big) Q_\alpha^0 + \frac{1}{\alpha}, 
 \enqs
 and plugging into \reff{interQ0}, we get the required result. 
 \ep

  \vspace{3mm}

By using  \reff{GQ} and \reff{Qinteg}, we then obtain the integral expression \reff{integG} of the function $G_\alpha$ as in Proposition \ref{propsignature}.  
Finally, we derive the asymptotic behavior of the mean signature plot.

\begin{Proposition}
Under {\bf (H)}, we get:
\beq
\bar V(\infty) \; := \;  \lim_{\tau\rightarrow\infty} \bar V(\tau) &=& \sigma^2_{_\infty}, \label{Vinfty} \\
\bar V(0^+) \; := \;  \lim_{\tau\downarrow 0^+} \bar V(\tau) &=&  \bar\lambda \E[\xi_n^2], \label{V0}
\enq
and 
\beqs
\bar V(0^+) > \bar V(\infty) & \mbox{ {\rm if and only if} } & \alpha < 0. 
\enqs
\end{Proposition}
{\bf Proof.}  By observing that the function $G_\alpha$ is  stricly bounded in $\tau$ by $1$ for any $\alpha$ $\in [-1,1)$, we easily obtain from the  expression \reff{Vsign} the limit for $\bar V(\tau)$ when $\tau$ goes to infinity.

\vspace{1mm}

\noindent   On the other hand, by substituting the integral formula \reff{integG} of the function $G_\alpha$  into the  expression \reff{Vsign} of the mean signature plot, we have:
\begin{eqnarray*}
\bar V(\tau) 
&=& \sigma_{_\infty}^2  -  \frac{2\alpha}{1-\alpha} \big(\mathbb{E}[\xi_n] \big)^2 
\Big(\frac{\bar\lambda}{\alpha}\Big) \Big(1 -\big(1-\alpha\big)\frac{1}{\tau} \int_0^\tau Q_\alpha^0(s)ds\Big)  
\end{eqnarray*}
Next, by noting that
\begin{eqnarray*}
\lim_{\tau\rightarrow 0^+} \frac{1}{\tau}\int_0^t Q_\alpha^0(s)ds &=&  Q_\alpha^0(0) \; = \;   1,   
\end{eqnarray*}
we deduce that
\beq \label{Vinfty0}
\lim_{\tau\rightarrow 0^+}  \bar V(\tau)  &=& \sigma_{_\infty}^2 - \frac{2\bar\lambda\alpha}{1-\alpha} \big(\mathbb{E}[\xi_n]\big)^2,
\enq
which gives \reff{V0} from the expression \reff{sigmamacro} of $\sigma_{_\infty}^2$.  Finally, we immediately see  from \reff{Vinfty0} that 
$\bar V(0^+) > \bar V(\infty)$ if and only if $\alpha$ $<$ $0$. 
\ep





\vspace{3mm}



\end{document}